\documentclass[twocolumn,showpacs,preprintnumbers,amsmath,amssymb]{revtex4}
\usepackage{tabularx,graphicx}\begin{document}
\newcommand{\beq}{\begin{equation}}
\newcommand{\eeq}{\end{equation}}
\newcommand{\beqn}{\begin{eqnarray}}
\newcommand{\eeqn}{\end{eqnarray}}
\newcommand{\bmath}{\begin{subequations}}
\newcommand{\emath}{\end{subequations}}
\def\skipline{\par \vspace{\baselineskip}}
\title{Superconductivity, diamagnetism, and the mean inner potential of solids
}
\author{J. E. Hirsch }
\address{Department of Physics, University of California, San Diego\\
La Jolla, CA 92093-0319}
 
\begin{abstract} 

The mean inner potential of a solid is known to be proportional to its diamagnetic susceptibility.
Superconductors exhibit giant diamagnetism. What does this say about the connection between superconductivity and mean inner potential? 
Nothing, according to the conventional theory of superconductivity. Instead, it is proposed that   a deep connection exists between the mean inner potential, 
diamagnetism, and superconductivity: that they are all intimately linked  to
 the fundamental charge asymmetry of matter.
It is  discussed how this physics can be
probed  experimentally and what  the implications of  different experimental findings would be for
the understanding of superconductivity.
\end{abstract}
\maketitle

\section{introduction}
To explain quantitatively the Davisson-Germer electron diffraction experiments\cite{dg} that established the wave nature of the electron in 1927, it was necessary to assume that either
the spacing  between atoms near the surface was different than in the bulk\cite{dg,pat}, or that the electron wavelength was different in vacuum than
inside the crystal\cite{eck}. Bethe\cite{b0,b} adopted the second point of view (which turned out to be the correct one), 
pointing out that Schr\"odinger's equation naturally leads to a different electron wavelength in the interior of the crystal
due to the change in its potential energy, and initially proposed\cite{b0} that the electron wavelength becomes $longer$ inside the crystal, corresponding to a 
$positive$ potential energy, to explain the assignment of diffraction peaks made by Davisson and Germer\cite{dg}. However, soon thereafter Bethe corrected himself\cite{b}
coming to the conclusion that a different assignment of diffraction peaks was necessary, corresponding to a $negative$ potential energy for the electron
inside the crystal and a shorter electron wavelength. Otherwise\cite{b} ``m\"ussten die Leitungselektronen den Kristall spontan verlassen''  (``the 
conduction electrons would have to
spontaneously leave the crystal'') which obviously is not
the case.

Thus, even to Bethe it wasn't immediately obvious that the ``mean inner potential'' $V_0$,  the average electric potential sensed by the electron beam scattering off the
 planes in the interior of the crystal, is   positive. It became very clear to him however in his seminal 1928 paper\cite{b}, where he stated that
very generally ``$V_0$ ist also notwendig positiv'' (``$V_0$ is therefore necessarily positive''), and 
``Die Elektronen werden beim Eintritt in den Kristall beschleunigt''   (``The electrons are accelerated when entering the crystal''), contrary to the statements in his
earlier paper\cite{b0}.

The mean inner potential of a solid is necessarily positive because it reflects the fundamental charge asymmetry of matter, namely the fact  that the electron is lighter than
the proton. For example, in a hypothetical solid composed of electrons and positrons, e.g. two interpenetrating Wigner crystals of charges of opposite sign and equal mass, the
mean inner potential would  be zero by symmetry. Because quantum mechanics links the spatial extent of a charged particle's wavefunction and the associated charge distribution 
to the particle's mass, a solid composed of oppositely charged particles that have $not$ the same mass will have a mean inner potential that is $not$ zero. 
The sign of the mean inner potential is the same as the sign of the $heavier$ particle charge. Charged particles entering the crystal 
(as in Davisson Germer's experiment) will
be accelerated (deccelerated) if the sign of their charge is the same as
(opposite to)  that of the $lighter$ particles in the crystal.

Soon after Bethe's paper it was pointed out by Rosenfeld\cite{rosen} and others\cite{pauling,slater} that the mean inner potential of a solid is intimately linked to its
diamagnetic susceptibility. This is because both depend in the same way on the spatial extent of the electronic
wavefunctions in the solid\cite{miyake,radi,spencek}. Note however that diamagnetism is $not$
sensitive to the $sign$ of the charges involved, nor is it conditional on charge asymmetry: a solid with antiprotons in its nuclei and positrons orbiting around them would have exactly the same diamagnetic 
susceptibility as its real world counterpart, and it would remain diamagnetic even in the limit where positive and negative particles have the same mass. Nevertheless, once charge symmetry
$is$ broken, breaking it even more   e.g. by  making the mass of the lighter particle even lighter
will increase the magnitudes of both the magnetic susceptibility and the mean inner potential.

The theory of hole superconductivity\cite{holesc} proposes that charge asymmetry is at the root of the phenomenon of superconductivity\cite{ehasym}. This is suggested for example by the
observation that superconductors tend to have positive Hall coefficients in the normal state\cite{chapnik}. More to the point, the fact that a rotating
superconductor acquires a magnetic moment (``London moment''\cite{lm}) and as a consequence generates a magnetic field that is always
$parallel$, never $antiparallel$ to its angular velocity, reveals a deep link between charge asymmetry and
superconductivity\cite{rotating}. We have argued that the Meissner effect is also a reflection of the fundamental charge asymmetry of matter\cite{meissner},
originating in expulsion of {\it negative charge} from the interior to the surface of the superconductor. However, in the Meissner effect
the charge asymmetry is hidden, since the Meissner effect does not reveal the $sign$ of the charge carriers involved in the Meissner current as the London moment does\cite{rotating}.
We propose in this paper that the fact that the Meissner effect is linked to diamagnetism, and diamagnetism is linked to the necessarily $positive$ mean
inner potential of solids, reveals the  deep link between the Meissner effect (and hence superconductivity) and charge asymmetry in nature.

This proposal is at odds with the conventional understanding of superconductivity as described by London\cite{london} and BCS\cite{bcs} as well as by further
developments of the conventional theory\cite{parks}. Within the conventional understanding charge asymmetry plays $no$ role
in superconductivity. BCS theory in its simplest form is electron-hole symmetric,
the condensate carries no net charge and quasiparticle excitations are charge neutral on average. 
Thus, there is a fundamental conflict between the conventional understanding of superconductivity and what we propose is the correct understanding of 
superconductivity. 
We argue that within the conventional theory there is no explanation for the universal sign of the
London moment  nor for the predominance of positive Hall coefficients among superconductors, nor is there an explanation
of the origin of the Meissner effect\cite{validity}.

In recent work\cite{holog1,holog2} we calculated the expected increase in the mean inner
potential resulting from the internal electric field that develops when negative charge is expelled
from the interior to the surface of the superconductor. This effect is predicted to be measurable
only at temperatures much lower than $T_c$, because at higher temperatures it is screened
by thermally excited normal quasiparticles\cite{holog1,holog2}. In this paper however we find that the connection
between mean inner potential and superconductivity is much deeper than we had
originally realized. Charge asymmetry, diamagnetic susceptibility,
kinetic energy lowering, mean inner potential and superconductivity are 
inextricably linked in our theory. The key new experimental
prediction that results from this realization is that an increase in the mean inner potential
should set in immediately when cooling a material below $T_c$,   with the
effect discussed in our earlier work\cite{holog1,holog2} giving an additional increase at much lower temperatures.

In section II   we discuss the connection between mean inner potential, diamagnetic susceptibility and charge asymmetry in solids.
In Sect. III we discuss dynamic Hubbard models\cite{dynhub} that we have   proposed to describe charge asymmetry in solids\cite{holeelec} and to 
explain superconductivity\cite{hm}
within the theory of hole superconductivity. Sect. IV discusses the 
calculation of the mean inner potential within a dynamic Hubbard model, and in Sect. V we discuss  the expected changes in the mean inner potential
as a function of temperature within this description. In Sects. VI-VIII we calculate the expected change in the mean inner potential assuming various possible
scenarios, and in Sect. IX   we examine the possibility to detect these effects experimentally.   We conclude in Sect. X with a discussion and summary.

\section{Mean inner potential, diamagnetic susceptibility and charge asymmetry}

The mean inner potential of a solid is defined by\cite{b}
\beq
V_0=\frac{1}{\Omega}\int d^3 r V(\vec{r})
\eeq
where $V(\vec{r})$ is the total electric potential at point $\vec{r}$ due to all the charges in the solid,  $\Omega$ is the volume of the solid and the integral is over
the entire crystal volume. Assuming a periodic structure with periodic boundary conditions, unit cells of volume $v$ and $n$ atoms in the unit cell, one can perform the integral
Eq. (1) using the charges in any one   unit cell,
\beq
V_0=\frac{1}{v}\sum_{i=1}^n \int  d^3r V_i(\vec{r})
\eeq
where $V_i$ is the electric potential due to the charge of atom $i$ in the unit cell, given by
\beq
V_i(\vec{r})=\int d^3r' \frac{\rho_+^i(\vec{r'})+\rho_-^i(\vec{r'})}{|\vec{r}-\vec{r'}|}
\eeq
with $\rho_+^i, \rho_-^i$ the densities of positive charge and negative charge from atom $i$. The integral in Eq. (3)  is over all space,
 the main contribution to it comes from
integration over the volume of the unit cell where the atom in question is located.

We assume, following Bethe, a spherically symmetric charge distribution centered at the atom location. Using the relation valid for any spherically symmetric function 
$f(\vec{r})=f(r)$
\beq
\int d^3r' \frac {f(\vec{r'})} {|\vec{r}-\vec{r'}|}=\frac{4\pi}{r}[\int_0^rdr' (r')^2 f(r')+\int_r^\infty dr' r r' f(r')]
\eeq
and assuming the atoms are charge-neutral, i.e. 
\beq
\int d^3r [\rho_+^i(\vec{r'})+\rho_-^i(\vec{r'})]=0
\eeq
Eq. (3) can be rewritten as
\beq
V_0=-\frac{2\pi}{3v}   \sum_{i=1}^n \int  d^3r   [\rho_+^i(\vec{r})+\rho_-^i(\vec{r})] r^2
\eeq
Therefore, the contribution of the charge density to the average potential is weighted more heavily the further away the charge is from the center,
and the sign of the contribution is $opposite$ to  the sign of the charge. As Bethe pointed out\cite{b},
$V_0$ is a kind of ``moment of inertia'' of the charge distribution, albeit with opposite sign.

On the other hand, the diamagnetic susceptibility of the solid can be written as
\beq
\chi_{dia}=-\frac{1}{6v c^2}\sum_{i=1}^n\int d^3 r  [\frac{q_i}{m_i}\rho_+^i(\vec{r})+\frac{e}{m_e}\rho_-^i(\vec{r})] r^2
\eeq
where $m_e$ and $e<0$ are the electron mass and charge, and  $m_i$ and $q_i>0$ are the mass and charge of the i-th  nucleus. Both terms in Eq. (7) have the same 
sign, so the susceptibility is sensitive to the spatial extent of the charge distribution as the mean inner potential is but 
in contrast to it is independent of the sign of the charge.

The Schr\"odinger equation determines the distribution of positive and negative charges in the solid, and because the positive charge (nucleus)  is much heavier than the negative
charge it is much more localized, in fact it is essentially a $\delta-$function. As a consequence, the integrals in Eqs. (6) and (7) are completely dominated by the contribution
of the lighter negative charge (electrons). Hence the integral in Eq. (6) is  negative, giving rise to a necessarily positive mean inner potential $V_0$, of opposite sign to the diamagnetic
susceptibility which is of course always negative.

Assuming the positive charge is all located at $r=0$ it does not contribute to either Eq. (6) or Eq. (7) and we have
\beq
V_0=-\frac{2\pi}{3v}   \sum_{i=1}^n \int  d^3r   \rho_-^i(\vec{r})  r^2=-\frac{2\pi e}{3v}   \sum_{j; i=1}^n <r_j^2>_i
\eeq
where $<r_j^2>_i$ is the average of $r^2$ for the $j-$th electron in atom $i$, so it increases when the size of the orbit increases.
The diamagnetic susceptibility is given by $V_0$ times a multiplicative constant
\beq
\chi_{dia}=- \frac{e}{6vm_e c^2}\sum_{i=1}^n\int d^3 r  \rho_-^i(\vec{r})r^2=
\frac{e}{4m_e c^2}V_0
\eeq
as first derived by Rosenfeld\cite{rosen}, who called it a ``strange relationship'' (``merkw\"urdige Beziehung''). Indeed, the full expressions Eq. (6) and Eq. (7) are not
proportional to each other since the positive and negative charges contribute with opposite signs   to Eq. (6) and with the same sign to Eq. (7). It is only because in nature
  the positive charge is $much$ heavier than the negative charge that the (approximate)  proportionality results. 
In a typical solid $V_0\sim$ few Volts and $\chi_{dia}\sim 10^{-6}$, consistent with Eq. (9).

In a finite crystal there will be contributions to the mean inner potential from the surface dipole layer resulting from some electronic charge
spilling out beyond the surface of the crystal. However, as  discussed by O'Keeffe and Spence\cite{spencek},
it is often the case that calculating the mean inner potential for an infinite crystal using neutral ``pseudo-atoms'', as originally done by Bethe and reviewed above, properly takes into account the surface dipole contributions arising in a finite crystal. In addition, bonding effects will somewhat modify the mean inner potential from the
``non-binding'' approximation described above\cite{spencek,radi,gaj2,sanchez}. We will not consider these effects here.

So the essential points we wish to underline are: the mean inner potential is proportional to the diamagnetic susceptibility
because the positive charge is so much heavier than the negative charge that its contribution to both quantities can be neglected. The mean inner potential is always positive because
 the lighter negative particle (electron) has a more extended wavefunction than the heavier positive particle (nucleus) because the electron pays a higher price in {\it kinetic energy} if
 its  wavefunction becomes too compressed.
Recall that the kinetic energy for a particle of mass $m$ is $\sim \hbar^2/(2mr^2)$, with $r$ the spatial extent of the wavefunction.
 In other words, the mean inner potential is positive because electrons, being much lighter than the nuclei, expand their wavefunction well beyond the spatial
extent of the positive nucleus in order to lower their kinetic energy. Within our theory,  in the superconductor  electrons expand their wavefunction even more than in
the normal state to lower their kinetic energy even more\cite{kinenergy}, and as a consequence the mean inner potential becomes even more positive. Expanding the 
electronic wavefunction also  leads
to  an increase in   the diamagnetic susceptibility.

However, note that a superconductor is a {\it perfect diamagnet}, hence 
\beq
\chi_{dia}=-\frac{1}{4\pi}
\eeq
which would imply from Eq. (9) that $V_0=m_e c^2/\pi |e|=1.6\times 10^6$ Volts, five orders of magnitude larger than in the normal state, a result that clearly doesn't make sense.
Therefore we conclude that Eq. (9) {\it does not hold} in the superconducting state.
Nevertheless, we argue that Eq. (9) strongly suggests that when a system goes superconducting there should be an increase in the mean 
inner potential together with the increase in the diamagnetic susceptibility. 

\section{dynamic Hubbard model}

To study the mean inner potential it is essential to properly describe the size of atoms. 
The size of an atom is determined by the interplay between the electron-ion interaction, the electron-electron interaction and the electron kinetic energy\cite{slater}. 
In a solid, the wavefunctions and energies
are determined by the atomic configurations and their modification
in the solid state environment.

The Hubbard model is the simplest model proposed to describe the physics of electrons in solids
including the effects of electron-electron interactions, and is being extensively used. We have argued that
the Hubbard model is fundamentally  flawed\cite{inapp}  because it ignores the effect of the electron-electron repulsion on 
the $state$ of the electrons in  the atom, and as a consequence it is  inadequate to describe the
physics of electrons in atoms and solids. Instead, we have proposed 
$dynamic$ Hubbard models\cite{dynhub} as the simplest class of models adequate for these purposes. 
These models describe the {\it orbital expansion} that takes place when a 
non-degenerate atomic orbital is occupied by two electrons, relative  to the
singly-occupied orbital. We have
furthermore argued that the distinguishing features of dynamic Hubbard models with respect
to the conventional Hubbard model are the key to understand superconductivity\cite{last,last2,holesc}.

 \begin{figure}
\resizebox{8.5cm}{!}{\includegraphics[width=7cm]{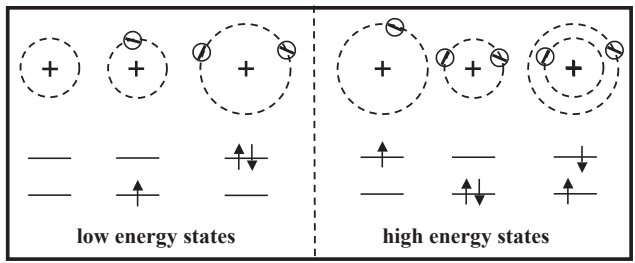}}
  \caption{ Different atomic states for zero, one and two electrons in the atom. The upper panels show the electronic orbits: the expanded orbit when two electrons are present
  gives lower energy. The lower  panels show the electronic states in a two-orbital model, with the
  upper (lower) orbital representing the expanded (unexpanded) orbit.   The three `low energy states' are the quasiparticle states in the effective low energy Hamiltonian
  resulting for this and similar models.}
  \end{figure}

Figure 1 shows schematically the atomic orbits of different sizes.
The lowest energy states correspond to an unexpanded  orbit when the orbital has one electron, and an
expanded orbit when the orbital has two electrons. The orbital expansion lowers the electron-electron repulsion energy as well as each electron's kinetic
energy, paying a (smaller) price in electron-ion energy\cite{ehole3}. High energy states result when the
orbits are different, as shown on the right side of Fig. 1: e.g. a non-expanded orbit when 
two electrons occupy the orbital or an expanded one when one electron occupies it. One way to model this physics   is with two electronic states per site 
as shown in Fig. 1, with appropriate
parameters for the single electron energy and the interactions\cite{twoorb}.  The Hamiltonian is 
\bmath
\beq
H=\sum_i H_i-\sum_{<ij>,\nu,\nu',\sigma} t_{\nu \nu'} (c_{i\nu\sigma}^{\dagger} c_{j \nu' \sigma}+h.c.)
\eeq
\beq
H_i=\epsilon n_2 - t'(c_{1 \sigma}^{ \dagger}c_{2 \sigma}+h.c.)+\sum_\nu U_\nu n_{\nu \uparrow} n_{\nu \downarrow}
+Vn_1 n_2
\eeq
\emath
where $\nu=1,2$ label the lower and upper orbitals, $n_\nu=\sum_\sigma n_{\nu\sigma}$ and we have omitted the site index on the right side of Eq. (11b).
$\epsilon$ denotes the difference in single particle energies between both levels, which results from a cost (increase) in
electron-ion energy partially compensated by a  gain (lowering) of kinetic energy. The conditions on the interaction parameters so that the lowest state 
of the doubly occupied orbital is with both electrons 
in the upper level, as shown in Fig. 1, are  $U_2+2\epsilon <U_1$, $U_2+\epsilon<V$.
Other ways to model this physics is with one electron state per site and an auxiliary boson degree
of freedom, which can be a harmonic 
oscillator\cite{dynhub}   or a spin 1/2 degree of freedom\cite{spin}.

The low energy effective Hamiltonian for these models is\cite{holeelec} 
\bmath
\beq
H_{eff}=-\sum_{ij\sigma}  t_{ij}^\sigma [ {c}_{i\sigma}^\dagger  {c}_{j\sigma}+h.c.]+U_{eff} \sum_i  {n}_{i\uparrow} {n}_{i\downarrow}
\eeq
\beq
t_{ij}^\sigma=t [1+(S-1) n_{i,-\sigma} ][1+(S-1)n_{j,-\sigma} ] 
\eeq
\emath
where the factor $S$ is the overlap matrix element between the 
unexpanded and expanded atomic orbitals in Fig. 1. 
For a hydrogenic wavefunction
\bmath
\beq
\psi(r)=(\frac{Z^3}{\pi a_0^3})^{1/2} e^{-Zr/a_0}
\eeq
with $Z$ the charge of the ion when there are no electrons in the orbital under consideration. When the orbital is doubly occupied, the wavefunction for each electron 
expands to
 \beq
\bar{\psi}(r)=(\frac{\bar{Z}^3}{\pi a_0^3})^{1/2} e^{-\bar{Z}r/a_0}
\eeq
\emath
within a Hartree approximation, with
\beq
\bar{Z}=Z-\frac{5}{16}
\eeq
or more generally, according to Slater's empirical rules\cite{slater}
\beq
\bar{Z}=Z-0.35
\eeq
for any electrons other than those in the $1s$ shell. The overlap matrix element between expanded and unexpanded orbitals is
\beq
S=<\bar{\psi}|\psi>=\frac{(Z\bar{Z})^{3/2}}{((Z+\bar{Z})/2)^3} .
\eeq
In the Hamiltonian Eq. (12), the hopping amplitudes when there are zero, one and two other electrons at the two sites involved in the hopping process are
$t$, $St$ and $S^2t$. The difference in the last two values is the correlated hopping term
$\Delta t=tS(1-S)$ that drives pairing and superconductivity within a BCS treatment of this model\cite{hm}.  
These phenomena are strongest for small $S$, which according to Eq. (16) corresponds to small values of the ionic charge $Z$.

\section{dynamic Hubbard model and mean inner potential}

Superconductivity involves pairing of electrons\cite{bcs}, and in the strong coupling limit the
spatial extent of the pairs is very small. Consider the cartoon picture of superconductivity
shown in Fig. 2. Two electrons in different atoms in the normal state occupy the same
atom in the superconducting state. What will be the effect of this
transition for the mean inner potential of the solid?

\begin{figure}
\resizebox{8.5cm}{!}{\includegraphics[width=7cm]{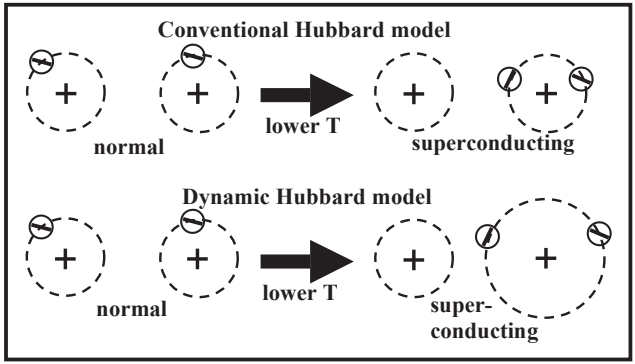}}
  \caption{ In a superconductor in the   strong coupling limit, electrons will pair on the same site.
  The upper panel of the picture shows this process within the conventional Hubbard model, 
  where the atomic orbitals are not modified by double occupancy, the lower
  panel shows it within the dynamic Hubbard model, that describes orbital expansion when an
  orbital becomes doubly occupied.}
\end{figure} 

Within the conventional Hubbard model   the orbital is not modified
by double occupancy and as a consequence the  mean inner potential will be unchanged.
The solid is now described by  a wavefunction that is a superposition of configurations with
doubly occupied and empty atomic orbitals in different spatial arrangements.
On average, the contribution to the mean inner potential of this wavefunction is
exactly the same as that of the neutral atoms shown in the left panels of Fig. 2.

Instead, the dynamic Hubbard model describes expansion of the doubly occupied orbital,
and the contribution to the  mean inner potential from such an ion is larger than
if the orbital did not expand. As a consequence, as pairing occurs and
the system goes superconducting the mean inner potential of the solid will increase.

It is easy to estimate the magnitude of the mean inner potential shift in this  strong coupling limit.
Assume the carrier concentration is $n_s$ carriers per unit volume, and that in the normal state
the orbitals are singly occupied and become doubly occupied in the superconducting state.
The mean inner potential can be obtained from Eq. (8). In the normal state, $\rho_-^i$ is the charge
density of one electron in the singly occupied orbital of the atom ($\equiv \rho_-^s$). In the superconducting state,
as indicated in Fig. 3, the volume per atom $v$ in Eq. (8) gets replaced by $2v$ and
the charge density is that of two electrons in the doubly occupied expanded orbital ($\equiv \rho_-^d$). The change in the mean inner
potential is then
\bmath
\beq
\Delta V_0=-\frac{2\pi}{3}n_s\int d^3r [\frac{1}{2}\rho_-^d(r)-\rho_-^s(r)]r^2
\eeq
or equivalently
\beq
\Delta V_0=-\frac{2\pi e}{3}n_s  [<r^2>_d-<r^2>_s]
\eeq
\emath
where $n_s$ is the superfluid density and $<r^2>_d$, $<r^2>_s$ are averages for a single electron for the atom with doubly and singly occupied
orbitals respectively (which are identical in the conventional Hubbard model).

  \begin{figure}
\resizebox{8.5cm}{!}{\includegraphics[width=7cm]{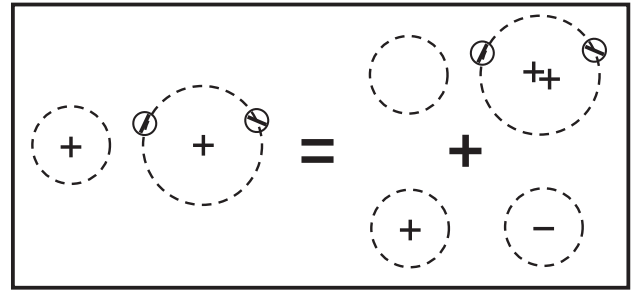}}
  \caption{ The electric potential from the two atoms on the left, one doubly occupied and one unoccupied, is
  equivalent to the sum of the potentials from the two  charge configurations on the right: a single doubly occupied
  orbital in an atom with an extra positive charge in the nucleus (upper right), plus a positive and a negative nucleus with
  no electrons (lower right). For the solid as a whole, the lower right configurations will give no contribution to the
  mean inner potential on average.}
\end{figure}

Using the wavefunctions Eq. (13) the charge densities in Eq. (17) are
\bmath
\beq
\rho_-^s(r)=e|\psi(r)|^2
\eeq
\beq
\rho_-^d(r)=2e|\bar{\psi}(r)|^2 , 
\eeq
\emath
the average radial extents are
\bmath
\beq
<r^2>_s=\frac{3a_0^2}{Z^2}
\eeq
\beq
<r^2>_d=\frac{3a_0^2}{\bar{Z}^2}
\eeq
\emath
and the change in the mean inner potential upon pairing is  
\beq
\Delta V_0=2\pi e n_s a_0^2 [\frac{1}{Z^2}-\frac{1}{\bar{Z}^2}]        
\eeq
which is always $positive$, never negative, since atomic orbitals always $expand$ upon double occupancy, never contract.

Consider the particular case where the superfluid charge density is given by
\beq
n_s=\frac{1}{4\pi a_0^3}  .
\eeq
Eq. (20) is
\beq
\Delta V_0=\frac{e}{2a_0}    [\frac{1}{Z^2}-\frac{1}{\bar{Z}^2}]=
\frac{13.6 eV}{e} [\frac{1}{Z^2}-\frac{1}{\bar{Z}^2}]   .
\eeq
Thus, for example for $Z=1$, using Eq. (15)  
\beq
\Delta V_0=1.57 Volts
\eeq
The value of superfluid density given by Eq. (21) is interesting because for that value 
a remarkable scaling relation holds between three different length scales\cite{meissner}, the superconducting
length scale $2\lambda_L$, the atomic length scale $a_0=\hbar^2/m_e e^2$ and the electron 
length scale (quantum electron radius) $r_q=\hbar/2m_ec$. It corresponds to 
a London penetration depth
\bmath
\beq
\lambda_L=(\frac{m_e c^2}{4\pi n_s e^2})^{1/2}
\eeq
that takes the value
\beq
\lambda_L=\frac{a_0}{\alpha}=137a_0
\eeq
\emath
with $\alpha=e^2/\hbar c$ the fine structure constant, or, numerically, $\lambda_L=72.47 \AA$, which may be a lower limit for possible values of the London penetration depth. 
We have proposed  that the Meissner effect results from electrons in the superconducting state occupying orbits of radius $2\lambda_L$\cite{sm}, and 
long ago, Slater\cite{slater2} 
 and others\cite{other1,other2,other3}
pointed out that the Meissner effect can be understood if electrons in superconductors reside
in orbits of order $137$ atomic diameters. We can think of the value of the superfluid density
Eq. (21) as resulting from 
electronic wavefunctions of radius $a_0$, i.e. spheres of volume $\frac{4\pi}{3}a_0^3$, with  $2/3$ of the spheres occupied by electrons and  $1/3$ of the
spheres empty in the normal state, as shown in Fig. 4. Upon pairing, the situation reverses and there are $2/3$ empty spheres and $1/3$ spheres occupied by electrons. 
Note that this corresponds to $n_s$ being the density of $holes$ rather than of electrons. Within our theory superconductivity can only exists when the
band is more than half-filled with electrons, or less than half-filled with holes, and the superfluid density that determines the London penetration depth is the 
density of holes, going to zero as the band becomes full with electrons\cite{london92}. The case considered here corresponds to a density of holes
per site $n_h=0.66$. This 
we suggest may be  the   `canonical case' of  superconductivity, i.e. the simplest situation
that displays the essence of superconductivity, akin to the hydrogen atom in atomic physics.

  \begin{figure}
\resizebox{8.5cm}{!}{\includegraphics[width=7cm]{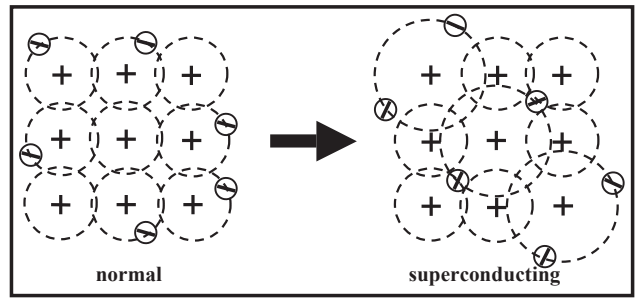}}
  \caption{ Canonical case, superfluid density given by Eq. (18) and $Z=1$. The radius of the singly occupied orbit is $a_0$ and the volume per atom is $(4\pi/3) a_0^3$. In the normal state,
  $2/3$ of the orbitals are singly occupied, $1/3$ are empty. In the superconducting state, $2/3$ of the orbitals are empty and $1/3$ are doubly occupied,
  with $a_0/(1-5/16)$  the radius of the orbit of the doubly occupied orbital.}
\end{figure}

The change in the mean inner potential given by Eq. (23) is easily measurable by a variety of techniques. It can even be   larger than Eq. (23)
if a smaller value for the ionic charge $Z$ is assumed. On the other hand it assumes completely local pairing, i.e. vanishing Cooper pair size,
and a rather high superfluid density.
In the next sections we consider more realistic estimates. However, the extreme simple case that we considered in this section is useful to 
illustrate the essential physics.

\section{mean inner potential changes in superconductors}

Within our theory there are several different reasons for why the mean inner potential (mip) of a solid should change as the temperature is lowered and 
the solid  becomes superconducting, and possibly even at higher
temperatures before it becomes superconducting. We list them here and discuss each of them briefly, and elaborate on the most important ones in the following sections.

\skipline
\noindent {\bf (a) Increase in site double occupancy upon pairing:} Within a BCS description of the superconducting state there will be a slight increase in the probability of double occupancy of an orbital in the superconducting state.
As discussed in the previous section, this will increase the magnitude of the mean inner potential.
We will discuss this quantitatively in the next section.

  \skipline
\noindent {\bf (b) Negative charge expulsion and macroscopic electric field:}
Within our theory electrons are expelled from the interior of the metal towards the surface as the metal enters the superconducting state\cite{chargeexp}, resulting in an interior
electric field pointing towards the surface, as shown qualitatively in Fig. 5. As discussed in refs. \cite{holog1,holog2} this will give rise to an increase in the mean inner potential at sufficiently
low temperatures where the electric field resulting from this  redistribution of the superfluid density is not screened by normal quasiparticles.

  \begin{figure}
\resizebox{5.5cm}{!}{\includegraphics[width=7cm]{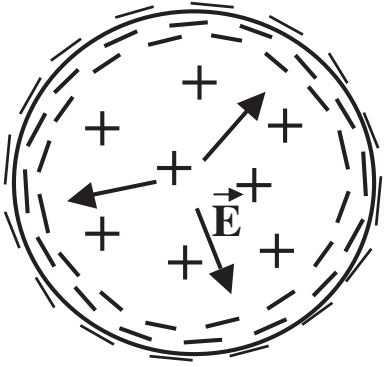}}
  \caption{ Charge distribution in the ground state of superconductors within the theory of hole
  superconductivity\cite{chargeexp}. There is excess negative charge within a London penetration depth of the surface, and negative charge spills out beyond the
  surface of the superconductor.}
\end{figure} 

  \skipline
\noindent {\bf (c) Spill-out effect:}
Our  theory predicts that some electrons spill out of the superconductor when charge expulsion occurs\cite{giantatom,electrospin}, as shown
qualitatively in Fig. 5. This will create a dipole layer at the
surface that will increase the mean inner potential in the interior, as discussed e.g. by O'Keefe and Spence\cite{spencek}.

  \skipline
\noindent {\bf (d) Orbit expansion:}
The theory predicts that electrons expand their orbits from radius $k_F^{-1}$ in the normal state ($k_F=$Fermi wavevector) to radius $2\lambda_L$
in the superconducting state\cite{sm}. This provides a dynamical explanation of the Meissner effect\cite{meissner}.
According to Eq. (8), this will lead to an increase in the mean inner potential (and in the diamagnetic susceptibility). Because the $2\lambda_L$ orbits are highly
overlapping however, the effect is not nearly as large as a straightforward application of Eq. (8) would indicate.

  \skipline
\noindent {\bf (e) Superconductor as giant atom:}
As shown by Eq. (8) and emphasized by Spence\cite{spence2}, the mean inner potential is proportional to the ``size'' of the atoms in the solid.
According to our theory a superconductor is a ``giant atom'' both with respect to its electric and its magnetic properties\cite{giantatom}, hence one would naturally expect
a larger mean inner potential in the superconducting compared to the normal state.

  \skipline
\noindent {\bf (f) Non-conventional orbital occupation:}
We have recently suggested the possibility that the propagation of the electrons that are part of the superfluid
occurs through the expanded orbitals only, even when the orbitals are singly occupied\cite{ehole3}.  If this is correct, it will cause a substantial increase in the mean inner potential in the superconducting state,
with the increase starting immediately upon cooling below $T_c$. We discuss this further in Sect. VII.

 \skipline
\noindent {\bf (g) Negative charge expulsion in the normal state:}
We have recently shown\cite{last} that the dynamic Hubbard model has a tendency to expel electrons from the interior to the surface and to regions of local inhomogeneity
(e.g. grain boundaries) even in the normal state. Even though no macroscopic electric field can exist in the normal state,
it is likely that this physics will cause an increase
 in the mean inner potential of a strong coupling superconductor due to $local$ negative charge expulsion as the temperature is lowered while still
above $T_c$, for example in the pseudogap phase of cuprate superconductors.

 \skipline
Note that $all$ the   reasons listed above predict an $increase$ in the mean inner potential of the solid as the temperature is lowered,
and in particular when the solid enters the superconducting state. However, not all of these will give sufficiently large contributions that can be
detected with current experimental techniques, as discussed in the next sections.

\section{change in mip upon pairing within BCS theory}

Consider a system of $N$ atoms with unit cell volume $v$ and one atom per unit cell, that has $N_s$ atoms with singly occupied orbitals and $N_d$ atoms with
doubly occupied orbitals. The mean inner potential is
\beq
V_0=-\frac{2\pi}{3v}e[\frac{N_s}{N}<r^2>_s+2\frac{N_d}{N}<r^2>_d]
\eeq
Assume the system has 
carrier density $n_h$ holes per site and carrier density per unit volume $n_s$, hence
\beq
n_s=\frac{n_h}{v} .
\eeq
 The average number of doubly and singly occupied sites is
 \bmath
 \beq
 N_d=<n_{i\uparrow}n_{i\downarrow}> N
 \eeq
 \beq
 N_s=(n_h-2<n_{i\uparrow}n_{i\downarrow}>) N
 \eeq
 \emath
and the mean inner potential is given by
\beqn
V_0=&-&\frac{2\pi}{3}en_s[<r^2>_s+ \nonumber \\
& &2\frac{<n_{i\uparrow}n_{i\downarrow}>}{n_h} ( <r^2>_d-<r^2>_s)]
\eeqn
Within a BCS treatment we have
\bmath
\beqn
<n_{i\uparrow}n_{i\downarrow}>&=&<n_{i\uparrow}><n_{i\downarrow}>+<c_{i\uparrow}^\dagger c_{i\downarrow}^\dagger><c_{i\downarrow} c_{i\uparrow}> \nonumber\\
&=&\frac{n_h^2}{4}+f_0^2
\eeqn
with $f_0$ the anomalous on-site pair amplitude (assumed real)
\beq
f_0=<c_{i\downarrow} c_{i\uparrow}>=<c_{i\uparrow}^\dagger c_{i\downarrow}^\dagger>
\eeq
\emath
and the change in the mean inner potential as the system goes superconducting is
\beq
\Delta V_0=V_0^s-V_0^n=-\frac{4\pi}{3}e\frac{n_s}{n_h}[<r^2>_d-<r^2>_s]f_0^2
\eeq
or, using Eqs. (19) and (24a)
\beq
\Delta V_0=-\frac{e}{a_0} \frac{1}{n_h} (\frac{a_0}{\alpha \lambda_L})^2  [\frac{1}{\bar{Z}}^2-\frac{1}{Z^2}]f_0^2
\eeq

The low energy effective Hamiltonian Eq. (12) gives rise to pairing and superconductivity when
the band is close to full\cite{hm}. For the parameter regime appropriate to describe cuprate
superconductors, the maximum $T_c$ occurs for approximately $n_h\sim 0.05$.
In Ref. \cite{undr2} (Fig. 3) we plotted the anomalous on-site pair amplitude $f_0$ as a function of
$n_h$ for one case of plausible parameters in the Hamiltonian, and the maximum
value of $f_0$ was  $f_0\sim 0.04$. With a value of the London penetration
depth $\lambda_L=1400 \AA$ appropriate for high $T_c$ cuprates Eq. (31) then yields
\beq
\Delta V_0\sim2.2\times 10^{-3}V [\frac{1}{\bar{Z}^2}-\frac{1}{Z^2}]
\eeq
as a typical value of the shift in the mean inner potential of high $T_c$ superconductors resulting
from this effect. Clearly this is too small to be detected experimentally. 

In the underdoped regime of the cuprates the system approaches the strong coupling limit where the pair size can become a single lattice spacing\cite{strong}.
In that regime we find that the mean inner potential change could become up to a factor of $\sim 3$ larger than Eq. (32), which unfortunately is still 
too small to be detected.
For other
superconductors with lower transition temperatures the size of the Cooper pairs will be larger and the values of $\Delta V_0$ from
Eq. (31) will be even smaller.

In summary: within the conventional BCS theory of superconductivity and
models such as the conventional Hubbard model that do not take into account
orbital expansion when sites are doubly occupied, $no$ change in the mean
inner potential is expected upon onset of pairing and superconductivity. Within BCS theory of the
dynamic Hubbard model an increase in the mean inner potential originating in the site double occupancy increase that occurs 
in the superconducting state $is$ predicted, however
the magnitude of the mean inner potential change due to this effect is likely to be too small to be detected experimentally.

\section{non-conventional orbital occupation and consequences for the mip}

In this section we discuss the possibility that the BCS treatment of the low energy effective Hamiltonian for dynamic Hubbard models discussed in the previous section 
does not in fact get the physics right,  and suggest that a correct treatment of the physics may lead to much larger changes in the mean inner potential upon onset of superconductivity.

The effective Hamiltonian Eq. (12) follows from the Hamiltonian Eq. (11) and from other dynamic Hubbard model Hamiltonians describing the same physics
 assuming that when an electron leaves a doubly occupied expanded orbital the second electron instantaneously relaxes to the lowest energy state of the
 atom with the single electron. In the language of small polaron models\cite{holstein}, it corresponds to the case where the electrons carry the distortion of the background with them,
 corresponding to `diagonal transitions'. There is however the other limiting case, corresponding to `vertical transitions',
 where a change in the occupation of the site  leaves the 
background degree of freedom unchanged. For the atom, it means that when one electron leaves a doubly occupied orbital the second electron remains in the expanded orbital rather than relaxing to
the contracted orbital corresponding to single occupation. For the two-orbital model, it means that the propagation of the superfluid electrons only involves
the upper orbital at each site.

\begin{figure}  
\resizebox{7.5cm}{!}{\includegraphics[width=7cm]{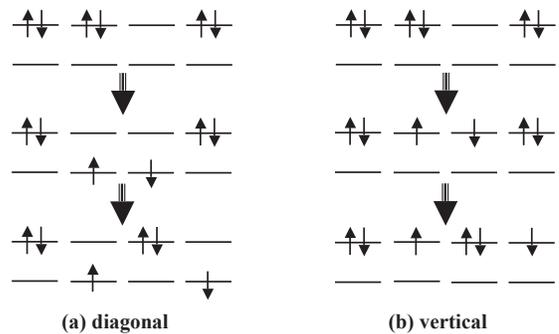}}
  \caption{ Motion of a pair of holes in the two orbital model. (a) denotes the   case considered previously, 
  diagonal transitions  in the equivalent small polaron model. (b) is the new proposal, corresponding to vertical transitions:
  the occupation of orbitals occupied by superfluid carriers is different to the conventional one shown in (a).}
\end{figure}

Figure 6 shows the propagation of a pair of holes with the equivalent two-orbital model. (a) corresponds to the   limit  
described by the effective Hamiltonian
Eq. (12), where only diagonal transitions are allowed. (b) shows the opposite limit  where only vertical transitions are allowed: the electron stays at the expanded orbital
 even when the site is singly occupied. 
Only the state depicted at the top of Fig. 6, where the 2 holes occupy the same site, is the same for both cases. 
We argue that there are   compelling physical reasons that follow from our earlier work on this problem for 
why the correct physics of the superfluid is described by Fig. 6(b) rather than Fig. 6(a), that we will discuss in the next section. First we discuss the consequences
of this new scenario for the mean inner potential.

Assuming that all the holes in the superfluid occupy expanded orbitals the mean inner potential in the superconducting state is
\beq
V_0^s=-\frac{2\pi}{3}en_s<r^2>_d
\eeq
In the normal state we have from Eq. (28) using  $<n_{i\uparrow} n_{i\downarrow}>= n_h^2/4$,
\beqn
V_0^n=&-&\frac{2\pi}{3}en_s[<r^2>_s+ \nonumber \\
& &\frac{n_h}{2} ( <r^2>_d-<r^2>_s)]
\eeqn
We can neglect the second term in Eq. (34) for small hole concentration, and the change in the mean inner potential is simply
\beqn
\Delta V_0&=&V_0^s-V_0^n=-\frac{2\pi}{3}en_s[<r^2>_d-<r^2>_s] \nonumber \\
&=&-2\pi e n_s a_0^2[\frac{1}{\bar{Z}^2}-\frac{1}{Z^2}]
\eeqn
or, using Eq. (24a) to express the superfluid density in terms of the London penetration depth
\beq
\Delta V_0=-\frac{e}{a_0} (\frac{a_0}{\alpha \lambda_L})^2 [\frac{1}{\bar{Z}^2}-\frac{1}{Z^2}]
\eeq
which is larger than the result obtained in the previous section (Eq. (31)) by a factor $n_h/f_0^2$. For $\lambda_L=1400\AA$,
\beq
\Delta V_0=0.037V\times  [\frac{1}{\bar{Z}^2}-\frac{1}{Z^2}]   .
\eeq
For the high $T_c$ cuprates, the orbital expansion should be appreciable since the effective nuclear charge of the oxygen negative ion $O^=$ is close to $0$.
Assuming for example $Z=0.6$ yields $r_d/r_s=Z/\bar{Z}\sim 2$ and $[1/\bar{Z}^2-1/Z^2]=11.1$, hence
\beq
\Delta V_0=0.41V  
\eeq
which is easily measurable. For conventional superconductors the expansion factor $r_d/r_s$ will be smaller but the London penetration depth will also be smaller
so one can expect appreciable values for $\Delta V_0$ for any superconductor if this scenario is valid.

\begin{figure}  
\resizebox{8.5cm}{!}{\includegraphics[width=7cm]{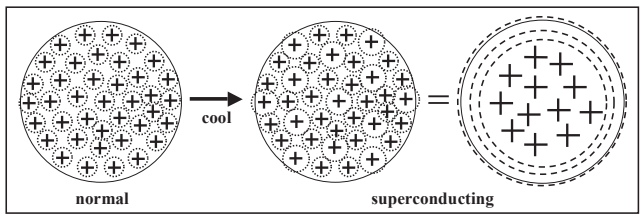}}
  \caption{In the non-conventional occupation scenario,
   as the system goes superconducting the orbitals where the superfluid carriers reside expand. This gives rise to a depletion of negative charge from the
  interior and extra negative charge near the surface, as well as negative charge spilling out beyond the surface of the body. }
\end{figure}

\section{arguments for non-conventional orbital occupation}

The scenario considered in the previous section assumes that the expanded orbitals of the doubly occupied atom remain expanded when one of the 
electrons
moves to another site. Clearly that is not the ground state configuration of the atom. However we propose that this is the
configuration that the singly occupied orbitals that are forming a Cooper pair and hence are part of the superfluid adopt. 
If so, when the superfluid density is $n_s$, a density $n_s$ of atoms will have expanded orbitals . This is shown qualitatively in Fig. 7.

First we show that this is plausible energetically. In the two-orbital model Eq. (11) the hopping amplitude for a single hole when there is orbital
relaxation (Fig. 6(a)) is $t_{12}S^2$, and if there is no orbital relaxation it is $t_{22}$. For a system with $z$  nearest neighbors to a site, a single hole
propagating through the lower orbital has energy
\bmath
\beq
E_1=-2zt_{12}S^2
\eeq
and one that propagates through the higher orbital has energy
\beq
E_1'=\epsilon-2zt_{22}
\eeq
\emath
so that the latter is favored provided
\beq
\epsilon < 2z (t_{22}-S^2t_{12})
\eeq
which is not implausible for systems where $S<<1$, even if $t_{12} \sim t_{22}$. Furthermore, since the higher orbital is more extended in space one has
in general $t_{22}>t_{12}$. Electrons propagating through the expanded orbitals pay a price in
electron-ion energy but lower their kinetic energy, both at the atom itself because of the
larger atomic radius and in the propagation between
atoms due to the larger hopping amplitude.

We believe  that  there are   compelling physical reasons that follow from our earlier work on this problem for 
why the propagation  {\it of the superfluid carriers}  involves only the expanded orbitals and no orbital relaxation as depicted in Fig. 6(b). These are:

\skipline

\noindent {\bf (a) Microscopic explanation of charge expulsion and spill-out:} the charge expulsion depicted in Fig. 5 is predicted by alternative macroscopic electrodynamic equations for superconductors that
we proposed in 2003\cite{chargeexp} . These equations are compelling for several reasons, as discussed by us\cite{chargeexp} as well as in London's original work\cite{londoneqs},
where similar equations were considered albeit later discarded\cite{londonexp}.  Our  formalism also predicts spill-out of some negative charge beyond the surface of the solid \cite{giantatom,electrospin}.
The non-conventional orbital occupation discussed here provides a $microscopic$ reason for the charge expulsion and spill-out:
if in the superconducting state   the atoms where single holes reside keep their orbital expanded, this will give rise to an overall expansion of the negative
charge distribution and consequent expulsion of negative charge from the interior to
the surface and beyond the surface.
\skipline
\noindent {\bf (b) Microscopic explanation of connection between kinetic energy driven superconductivity and negative charge expulsion:}
We have found in previous work that the condensation energy of the superconductor originates 
in  lowering of kinetic energy of the charge carriers (rather than of  potential energy as in
conventional BCS theory)\cite{kinetic,kinenergy}. The Hamiltonian predicting this, Eq. (12), is derived from the dynamic Hubbard model but does not explicitly describe the
atomic kinetic energy lowering resulting from orbit expansion. If in the superconducting state the singly occupied orbitals are expanded as 
depicted in Fig. 7, it shows that the kinetic energy lowering of carriers propagating in the lattice originates in  the $atomic$ kinetic energy lowering resulting from orbit expansion, which is also associated with
the negative charge expulsion. 

\skipline
\noindent {\bf (c)  Connection with diamagnetism:} diamagnetism increases when the size of the atomic electronic orbits increases, as given by Eq. (9). 
When cooling a metal into the superconducting state, diamagnetism starts to increase immediately below $T_c$. This suggests that electronic orbits expand
as the system is cooled below $T_c$ and the superfluid density increases, as depicted in Fig. 7.

        \begin{figure}  
\resizebox{8.5cm}{!}{\includegraphics[width=7cm]{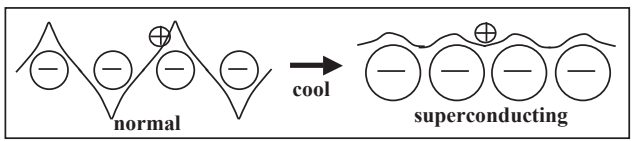}}
  \caption{In the normal state, carriers at the top of the band have a rapidly oscillating
  wave function and high kinetic energy, and the expanded orbitals shrink as
  the hole propagates through them, as shown on the left. In the superconducting state assuming the
  non-conventional orbital occupation scenario, carriers have a
  smooth wave function and low kinetic energy and propagate through the
  expanded orbitals without affecting them, as shown on the right.}
\end{figure} 

\skipline
 \noindent {\bf (d) Consistency with experiments showing complete `undressing':} We have argued in previous work\cite{ehasym2,holeelec2}  that experiments such as the 
 London moment, the gyromagnetic effect, the Bernoulli effect and the Hall effect below $T_c$ show that the superfluid carriers completely $undress$
 from the electron-electron and the electron-ion interaction. 
 In fact, this has also been proposed by others in early work on superconductivity\cite{rud,rud2}. As pointed out in Ref. \cite{holeelec2} that
 physics is $not$  
 described by the low-energy effective Hamiltonian Eq. (13) that only describes $partial$ undressing. Instead it corresponds to the superfluid carriers propagating
 without disturbing the background, which is consistent with propagation through expanded orbitals only as depicted in Fig. 6(b).
 
 \skipline
 \noindent {\bf (e) Consistency with wavelength expansion physics:} We have argued in previous work that superconductivity is associated with
 {\it wavelength expansion} of the carriers\cite{holeelec2}: in the normal state the wavelength of carriers is a single lattice spacing when the band is almost full, and the
 electrons interact strongly with the discrete ionic lattice thereby changing their effective mass from positive to negative; in the superconducting state the superfluid carriers expand their wavelength and no longer `see' the ionic periodicity but rather a uniform positive background\cite{giantatom}.
 This would be inconsistent with orbital relaxation occuring in one site or another as depicted in Fig. 6(a), but is consistent with the physics depicted
 in Fig. 6(b) where the superfluid carriers do not disrupt the state of other electrons.
 
 \skipline
    \noindent {\bf (f) Consistency with ``holes becoming electrons'':} We have argued in previous work\cite{ehole3} that in the superconducting state 
    holes migrate from the top to the bottom of the band adopting a smooth electron-like wavefunction, as shown  in Fig. 8. This corresponds to a 
    mirror-image of the electronic wavefunction for electrons at the bottom of the band, and is consistent
    with the atomic charge distribution remaining expanded when a hole lands on the site.
  
\skipline
        \noindent {\bf (g) Consistency with macroscopic phase coherence:}
        A superconductor has macroscopic phase coherence, resulting from a coherent superposition of Cooper pairs $c_{k\uparrow}^\dagger c_{-k\downarrow}^\dagger$.
        If the wavefunction of a member of a Cooper pair
        \beq
        c_{k\uparrow}^\dagger=\frac{1}{\sqrt{N}} \sum_i e^{i \vec{k}\cdot \vec{R}_i} c_{i\uparrow}^\dagger
        \eeq
        involves the lower electronic orbitals ($c_{i 1\uparrow}^\dagger$ in Eq. (11)) the wave function for
        $c_{k\uparrow}^\dagger c_{-k\downarrow}^\dagger$
         would have states where both electrons
        occupy the lower orbital rather the expanded one, that are high in energy and would have to be eliminated from the
        wave function, thus destroying the phase coherence.
        In other words, given  that the states shown at the top of Fig. 6 are part of the superfluid wavefunction, this  indicates that a phase-coherent wavefunction will 
        have other states in the wavefunction of the form shown on  Fig. 6 (b), with the single electron states in the same orbital as the two-electrons state, rather than
        the states shown in Fig. 6(a).
        
        \skipline
        \noindent {\bf (h) Consistency with the existence of a spin current in the ground state:} The lowest energy 
        state of superconductors within our theory in the absence of external fields
        has a macroscopic spin current flowing within a London penetration depth of the surface\cite{electrospin}, where carriers of opposite
        spin circulate in opposite directions without any interaction with the ions. This would not be compatible with propagation involving 
        orbital relaxation as in Fig. 6(a), but is further evidence that the superfluid carriers propagate without disrupting the background
        as depicted in Fig. 6(b).
        
          \skipline
        \noindent {\bf (i) Consistency with $^4He$  behavior:} 
        We have proposed that the physics of superconductors is closely related to the physics of
superfluid $^4He$ in that both phenomena are kinetic energy driven\cite{he1,he2}. As is well known, $^4He$ has negative thermal expansion coefficient below
$T_\lambda$, i.e. it  $expands$ as the temperature is lowered below the
superfluid transition temperature.
        The density decrease when $^4He$ is cooled starts to occur immediately below the
        transition into the superfluid state. Similarly  we expect the orbital expansion in the superconductor to start occurring immediately below $T_c$ due to
        the orbitals that are part of the superfluid expanding as depicted in Fig. 7, giving rise to an increase in the mean inner potential.

  \skipline

We argue that the various arguments discussed above
        provide compelling justification for a scenario where the atomic orbitals involved in the superfluid expand when the system is cooled below $T_c$,
        thus giving rise to an increase in the mean inner potential immediately after the system is cooled below $T_c$, as discussed in Sect. VII.
                A quantitative description of this physics, which has to take into account the effect of pair binding in preventing the orbital relaxation of the singly occupied site that is member
        of a Cooper pair,  is beyond the scope of this paper.  Whether or not this actually occurs can be determined experimentally as discussed in the next section.

\section{experimental detection}
The original Davisson-Germer experiment indicated the existence of a mean inner potential of about $13V$ in Ni by the fact that a change in wavelength
of the $54eV$ beam electrons from $1.670\AA$ in vacuum to $1.494\AA$ inside the material had to be assumed in order to explain the Bragg reflection angle
measured (see Bethe\cite{b}, table I). An additional  mean inner potential increase of $0.41V$ at lower temperatures, as given by Eq. (38), might even back then have been detectable as an additional wavelength reduction 
of $0.004\AA$. We believe that the various effects discussed in this paper associated with the predicted expansion of the electronic wavefunction and negative charge expulsion
 in the superconducting state should certainly be detectable with current experimental capabilities,
using low energy electron diffraction, reflection high energy electron difraction, in-line or off-axis electron holography\cite{gaj,saldin,yam,dunin}.

Assuming the non-conventional orbital occupation scenario discussed in the previous section is valid, it should be possible to detect the mean inner potential increase
as the sample is cooled below $T_c$ even if the beam electrons don't penetrate far into the sample. Thus one could perform a  low energy electron diffraction measurement as
done by Davisson and Germer, or a reflection high energy electron diffraction measurement with small incidence angle\cite{yam}. These experiments would detect a
small increase in the index of refraction of the material for electron waves
 beyond the value in the normal state. This effect should set in immediately as the sample is cooled below $T_c$ and increase in
magnitude following the magnitude of the superfluid density $n_s$ which is proportional to $\lambda_L^{-2}$ and rises linearly\cite{bcs} from zero at $T_c$.
The predicted spill-out of electrons beyond the surface and electric fields outside the sample near the surface\cite{ellipsoid} predicted by our theory
may be detectable by Fresnel contrast analysis\cite{dunin}.

The most direct and sensitive probe  would be   off-axis electron holography\cite{gaj,holog},  to
detect the predicted mean inner potential increase by detecting an increase in the phase shift of an  electron wave going through the
sample  relative to the wave going through vacuum. Here it would be best to use high energy electrons (e.g. $300 keV$ or higher) and as thick a sample as possible. The phase shift
for mean inner potential $V_0$ and a sample of thickness $d$ is given by 
\bmath
\beq
\varphi=C_TV_0 d ,
\eeq  with
\beq
C_T=\frac{ \alpha}{|e|} \frac{  (T+E_0)\sqrt{T^2+2TE_0}}{T(T+2E_0)}
\eeq
\emath
with   $T$ the electron kinetic energy, $E_0$ the electron rest energy and $\alpha$ the fine structure constant. 
Such experiments can easily detect a phase shift  as small as $2\pi/100$\cite{tonomura}, which for $300 keV$ electrons corresponds to a mean
inner potential increase of $0.10V$ for a $100 nm$ thick sample. Recent advances in accuracy have allowed phase observation with accuracy of
$2\pi/300$\cite{acc1} and even $2\pi/1500$\cite{acc2}.
In recent work\cite{holog1,holog2}  we have calculated and given quantitative estimates for the expected phase shift in electron holography
experiments resulting from the macroscopic charge redistribution and resulting macroscopic
electric field shown qualitatively in Fig. 5. That effect should only be detectable at sufficiently low temperatures that the macroscopic electric field is not
screened by thermally excited quasiparticles, which we estimated as approximately $T<0.16 T_c$, and only with a probe that penetrates the sample
beyond a London penetration depth of the surface.

        \begin{figure}  
\resizebox{8.5cm}{!}{\includegraphics[width=7cm]{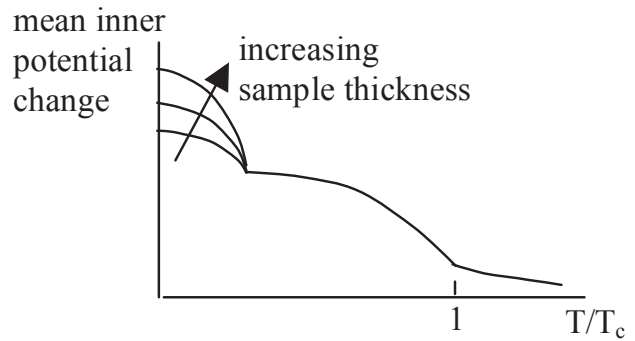}}
  \caption{Expected behavior of the mean inner potential versus temperature. The low temperature increase that depends on sample thickness occurs when the
  thermally excited quasiparticles no longer can screen the macroscopic electric field created by redistribution of the superfluid.}
\end{figure} 

Thus in an electron holography experiment we would expect, as shown in Fig. 9,  that as the sample is cooled below $T_c$ one would first detect the mean inner potential increase
due to the orbit enlargement of the superfluid carriers and negative charge spill-out beyond the surface, increasing proportionally to the superfluid density, and at
much lower temperatures one would see the additional increase due to the development of the macroscopic electric field
discussed in \cite{holog1,holog2}. The first shift in the mean inner
potential should be independent of sample thickness and detectable either by experiments probing the sample near the surface (i.e. in reflection experiments) or in the interior, 
while the second should increase with sample thickness and be detectable only for thick samples with the electron beam transmitted through the sample. 
Furthermore, as discussed in Sect. V, even in the normal state we expect an increase in the mean inner potential as the temperature is lowered in the strong coupling regime for
materials giving rise to
high $T_c$ superconductivity. e.g. in the underdoped `pseudogap' phase of high $T_c$ cuprate materials. 
The expected qualitative behavior of the mean inner potential versus temperature is shown
  in Fig. 9. 
  
  On the other hand, if the non-conventional orbital occupation scenario is not valid, we would expect no appreciable increase in the mean inner potential until the 
  sample is cooled sufficiently below $T_c$ that the macroscopic electric field shown in Fig. 5 starts to develop, $T<\sim0.16T_c$. 
Thus, what is observed experimentally will provide key information for further development of the theory. If $no$ additional phase shift is observed in the superconducting state at any
temperature below the superconducting transition temperature, it would provide very strong evidence $against$ the theory discussed here.

\section{summary and discussion}

The mean inner potential is a fundamental property of   solids and played a key role in the interpretation of the  seminal experiments that established the wave nature
of the electron\cite{dg}. In this paper we have suggested that the mean inner potential may play a key role in advancing our understanding of superconductivity, arguably the most
fundamental state of condensed matter since it displays quantum-mechanical behavior at the macroscopic scale. In turn, this could lead to a deeper understanding
of the electron itself\cite{electron,kinenergy}.

In summary: there are $qualitative$ reasons, independent of our theory of superconductivity,  that suggest that the mean inner potential of a metal should increase when the metal enters the superconducting state.
The mean inner potential is known to be proportional to the diamagnetic susceptibility, and the diamagnetic susceptibility increases dramatically in the
superconducting state. The mean inner potential is proportional to the ``size'' of the atoms, and a superconductor can be understood 
(in some sense) as a ``giant atom'', as remarked by many early workers in superconductivity\cite{slater2,londoneqs,london3,london4}, as well as  indicated by the fact
that a superconductor has macroscopic phase coherence. As pointed
out by Slater\cite{slater2} and other early workers\cite{other1,other2,other3},
the giant diamagnetism of superconductors can be understood if the electronic orbits in the superconducting state are of order $137$ lattice spacings, and the mean inner potential is proportional to the size of electronic orbits. Finally, superconductivity involves pairing, which should lead to an increase in the ``size'' of the atom when a pair resides on it, which would increase the mean inner potential.

Yet there has been no experimental report to date of any measured changes in the mean inner potential of materials becoming superconducting in
electron microscopy experiments\cite{s1,s2,s3,s4,s5}, nor has there been an experimental effort to look for
such changes. This is because the conventional BCS-London theory of superconductivity suggests {\it no connection} between the mean inner potential, diamagnetism,  and superconductivity and predicts
no change in the mean inner potential as a system becomes superconducting. This, we submit,
is disquieting. It suggests that either nature is
not as natural as it could have been,  or that there is something fundamentally wrong with the conventional theory of superconductivity.

Instead, we have proposed in this paper that there is a direct connection between the mean inner potential, diamagnetic susceptibility, and superconductivity:
that they are all intimately linked to the fundamental charge asymmetry of matter, namely the fact that the electron is $2000$ times lighter than the proton.
That the mean inner potential and the diamagnetic susceptibility are linked this way was discussed in Sect. II and we believe is incontrovertible.
That superconductivity is linked this way is a predicted consequence of  our alternative theory of superconductivity\cite{holesc} and will  certainly be controversial, given that essentially the entire physics community currently subscribes to the conventional theory of superconductivity
at least for the so-called ``conventional superconductors''.

As discussed in Sect. V, there are many different   reasons within our theory that indicate that the mean inner potential should increase in the superconducting
 state. They are all tied to the fact that superconductivity in our theory is {\it kinetic energy driven}, i.e. is associated with 
 electronic kinetic energy lowering\cite{kinenergy}, in contrast to the
 conventional theory that predicts electronic kinetic energy $increase$ in the superconducting state.   There is in fact some experimental evidence at least for
 high $T_c$ cuprates that superconductivity is kinetic energy driven\cite{kinexp}.
The reason that the mean inner potential and superconductivity are so closely linked within our theory is that the mean inner potential is also ``kinetic energy driven'':
 negative and positive charges separate in the atom and in the solid because the electron lowers its kinetic energy in this process despite the cost in
 potential (Coulomb) energy associated with charge separation\cite{emf}, 
 and this  gives rise to the (positive) mean inner potential. Further kinetic energy lowering as the system goes superconducting hence
 necessarily implies further  charge separation and further increase in the mean inner potential.  
 In an electron microscopy experiment this kinetic energy lowering that drives the mean inner potential, whether in the normal or in the superconducting state,  manifests itself in the $increased$ kinetic energy of the
 beam electrons going through the sample, as  first discussed by Bethe\cite{b}.

 Within the conventional theory of superconductivity, superconductivity results from pairing and this will in general  increase the probability of two electrons occupying
the same atomic orbital, hence 
taking into account the increase in the size of the orbit that necessarily results from double occupancy\cite{slater}  leads to an increase in the mean inner potential. However, the conventional theory of superconductivity ignores the orbital expansion that occurs when two electrons occupy the same orbital and thus, in its current form,  predicts no change in the mean inner potential. 
Once an increase in the mean inner potential in the superconducting state is measured experimentally there will surely be explanations proposed that would not fundamentally alter the conventional
theory. For example, it may be postulated that the nature of electronic states near the surface changes, or even the electronic configuration in the bulk changes, due to the opening
of the superconducting energy gap, in some particular way that can reproduce the experimental results. 
These explanations will not be compelling because they will not be linked to the Meissner effect, which supposedly is already
completely understood within the conventional theory\cite{bcs},  and   because they will be $post$dictive rather than $pre$dictive.

With respect to our theory, experimental results will have crucial implications. If an increase in the mean inner potential is detected $only$ at temperatures well below
$T_c$ (which would be surprising to us), it will negate the ``unconventional orbital occupation'' scenario discussed in Sects.  VII and VIII and support the theory in its early form\cite{hm}.
The low temperature increase in the mean inner potential will have to increase with sample thickness, otherwise it would contradict the charge expulsion and
electrodynamic equations predicted by our theory\cite{chargeexp, holog1,holog2}. If an increase in the mean inner potential is detected as soon as the
system is cooled below $T_c$, as we expect,  it will support and shed further light on the non-conventional orbital occupation scenario, and this 
will provide key input for further development of the theory. The expected additional increase in the mean inner potential at much lower temperatures resulting from
charge expulsion and macroscopic electric field should of course also be seen in that case. Finally, if no increase of the mean inner potential is seen in the
superconducting state at any temperature  it will falsify  the theory of hole superconductivity.

\acknowledgements
J.C.H. Spence's insightful work  on the subject of mean inner potentials and  discussions with him provided important stimulus for this paper, for which the author is grateful.
Stimulating discussions with R. Dunin-Borkowski, Yimei Zhu and H. Lichte are also gratefully acknowledged.


\begin{references} 
\bibitem{dg} C. Davisson and L.H. Germer, Phys. Rev. {\bf 30}, 705 (1927).
\bibitem{pat} A.L. Patterson, Nature {\bf 120}, 46 (1927).
\bibitem{eck} C. Eckart, PNAS {\bf 13}, 460 (1927).
\bibitem{b0} H. Bethe, Naturwissenschaften {\bf 15}, 786 (1927).
\bibitem{b} H. Bethe,  Ann. der Physik  {\bf 392}, 85 (1928).
 \bibitem{rosen} L. Rosenfeld, Naturwissenschaftern {\bf 17}, 49 (1929).
 \bibitem{pauling} L. Pauling, Phys. Rev. {\bf 34}, 954 (1929).
 \bibitem{slater} J.C. Slater, Phys. Rev. {\bf 36}, 57 (1930).
 
 \bibitem{miyake} S. Miyake, Proc. of the Phys.-Math. Soc. of Jpn {\bf 22}, 666 (1940).
 \bibitem{radi} G. Radi, Acta Cryst. A{\bf 26}, 41 (1969).
   \bibitem{spencek} M. O'Keefe and J.C.H. Spence, Acta Cryst. A{\bf50}, 33 (1994).
 
\bibitem{holesc} See references in http://physics.ucsd.edu/$\sim$jorge/hole.html.
\bibitem {ehasym} J.E. Hirsch,   J. Phys. Chem. Solids {\bf 67}, 21 (2006) and references therein.
\bibitem{chapnik} I.M. Chapnik, Sov. Phys. Dokl. {\bf 6}, 988 (1962).
\bibitem{lm} R. Becker, F. Sauter, and C. Heller, Z. Physik {\bf 85},
772, (1933).
\bibitem{rotating} J.E. Hirsch, Phys. Rev. B {\bf 68}, 012510 (2003). 
 \bibitem{meissner} J.E. Hirsch,   Physica Scripta {\bf 85},   035704 (2012).
 \bibitem{london} F. London, 
ÔSuperfluidsÕ , Volume I, Wiley, New York, (1950).
 \bibitem{bcs} M. Tinkham,  ``Introduction to Superconductivity'', 2nd ed, McGraw Hill, New York, 1996.
 \bibitem{parks} ÓSuperconductivityÓ, ed. by R.D. Parks, Marcel Dekker, New York, 1969.
 \bibitem{validity} J.E. Hirsch, Phys. Scripta {\bf 80},  035702 (2009). 
  \bibitem{holog1} J.E. Hirsch, Physica C {\bf 490}, 1 (2013).
  \bibitem{holog2} J.E. Hirsch, Ultramicroscopy {\bf 133}, 67  (2013).
 
  \bibitem{dynhub} J.E. Hirsch, Phys.Rev. Lett.  {\bf 87}, 206402 (2001).
  \bibitem{holeelec} J.E. Hirsch, Phys.Rev. B {\bf 65}, 184502 (2002).
    \bibitem{hm} J.E. Hirsch and F. Marsiglio, Phys. Rev. B {\bf 39}, 11515 (1989).


\bibitem{gaj2} M. Gajdardziska-Josifovska and A.H. Carim, in ``Introduction to Electron Holography'', ed. by E. V\"olkl, L.F. Allard and D.C. Joy, Kluwer, New York, 1999,
p. 267.
\bibitem{sanchez} A. Sanchez and M.A. Ochando, J. Phys. C{\bf 18}, 33 (1985).

   \bibitem{kinenergy} J.E. Hirsch, Int. J. Mod. Phys.  {\bf 25}, 1173 (2011).

   \bibitem{inapp} J.E. Hirsch, Physica B {\bf 199\&200}, 366 (1994).
 \bibitem{last} J.E. Hirsch,  Phys. Rev. B {\bf 87},  184506 (2013).
 \bibitem{last2} J.E. Hirsch,  Physica Scripta {\bf 88},   035704  (2013).
     \bibitem{ehole3} J.E. Hirsch, Int. J. Mod. Phys.  {\bf 23}, 3035 (2009).

     \bibitem{twoorb} J.E. Hirsch, Phys. Rev. B {\bf 43}, 11400 (1991); Phys.Rev. B {\bf 67}, 035103 (2003).
      \bibitem{spin} J.E. Hirsch, Phys. Lett. A {\bf 134}, 451 (1989).
 
    \bibitem{sm} J.E. Hirsch, Europhys. Lett. {\bf 81}, 67003 (2008).
 \bibitem{slater2} J.C. Slater, Phys.Rev. {\bf 52}, 214 (1937).
 \bibitem{other1} J. Frenkel, Phys. Rev. {\bf 43}, 907 (1933).
  \bibitem{other2} H.G. Smith, University of Toronto Studies, Low Temp. Series Vol. {\bf 76}, p. 23 (1935).
  \bibitem{other3}  V. Rudnitzky, Jour. Exp. Theor. Phys. {\bf 11}, 107 (1941).
 
 
 
 \bibitem{london92} J.E. Hirsch and F. Marsiglio,  Phys. Rev. B {\bf 48}, 4807 (1992).
  \bibitem{chargeexp} J.E. Hirsch, Phys.Rev. B {\bf 68}, 184502 (2003); Phys.Rev. B {\bf 69}, 214515 (2004).

  \bibitem{giantatom} J.E. Hirsch, Phys. Lett. A {\bf 309}, 457 (2003).
  \bibitem{electrospin}  J.E. Hirsch,   Ann. Phys. (Berlin)   {\bf 17}, 380 (2008).
  
\bibitem{spence2} J.C.H. Spence, Acta Cryst. A{\bf 49}, 231 (1993) and references therein.



\bibitem{undr2}  J.E. Hirsch,   Phys, Rev, B {\bf 62}, 14498 (2000).
\bibitem{strong}   F. Marsiglio and J.E. Hirsch, Physica C {\bf 165}, 71 (1990). 

\bibitem{holstein} T. Holstein, Ann. of Phys. {\bf 8}, 343 (1959); G.B. Arnold and T. Holstein, Ann. of Phys. {\bf 132}, 163 (1981).
 \bibitem{londoneqs} F. London and H. London, Proc. Roy. Soc. A{\bf 149}, 71 (1935);
Physica {\bf 2}, 341 (1935).

\bibitem{londonexp} H. London, Proc.Roy.Soc. A{\bf 155}, 102 (1936).
 \bibitem{kinetic}  J.E. Hirsch and F. Marsiglio, Phys. Rev. B {\bf 62}, 15131 (2000).
 
  \bibitem{ehasym2} J.E. Hirsch, Int. J. Mod. Phys.  {\bf 17}, 3236 (2003).
 \bibitem{holeelec2}  J.E. Hirsch, Phys.Rev. B{\bf 71}, 104522 (2005).
 

 
  \bibitem{rud2}  R.  Kronig, Zeit. f\"ur  Physik {\bf 78}, 744 (1932).
 \bibitem{rud} J. Frenkel and V. Rudnitzky, J. Exp. Theor. Phys. {\bf 9}, 260 (1939).

  \bibitem{he1} J.E. Hirsch, Mod. Phys. Lett. B {\bf 25}, 2219 (2011). 
\bibitem{he2} J.E. Hirsch,   Physica C    dx.doi.org/10.1016/j.physc.2013.03.010 (2013).

   \bibitem{gaj} M. Gajdardziska-Josifovska et al, Ultramicroscopy {\bf 50}, 285 (1993).

\bibitem{saldin}  D.K. Saldin and J.C.H. Spence, Ultramicroscopy {\bf 55}, 397 (1994).

 \bibitem{yam} N. Yamamoto and J.C.H. Spence, Thin Solid Films {\bf 104}, 43 (1983).
 \bibitem{dunin} R. E. Dunin-Borkowski, Ultramicroscopy {\bf 83}, 193  (2000).
 
 \bibitem{ellipsoid} J.E. Hirsch, Phys.Rev. Lett.  {\bf 92},  016402 (2004).
 
\bibitem{holog} A. Tonomura, ÒElectron HolographyÓ, Springer, Berlin, 1999.
\bibitem{tonomura} A. Tonomura et al,  Phys. Rev. Lett. {\bf 54}, 60 (1985).

\bibitem{acc1} K. Yamamoto et al, Jour. of Electron Microscopy {\bf 49}, 31 (2000).
\bibitem{acc2} T. Suzuki  et al, Ultramicroscopy {\bf 118}, 21  (2012).

 \bibitem{london3} H.G. Smith  and J. O. Wilhelm,  Rev. Mod. Phys.  {\bf 7},  232  (1935).
 \bibitem{london4} V. Ginsburg, Fortschritte der Physik {\bf 1}, 101 (1953).
 
\bibitem{electron} ``The Electron'', ed. by D. Hestenes and A. Weingartshofer, Kluwer Acad. Pub., Dordrecht, 1991.

\bibitem{s1} J.E. Bonevich et al, Phys. Rev. Lett. {\bf 70}, 2952 (1993).
\bibitem{s2}  K. Harada et al, Science {\bf 274}, 1167 (1996).
\bibitem{s3} J.E. Bonevich et al, Phys. Rev. B{\bf 57}, 1200 (1998).

\bibitem{s4}  A. Tonomura et al, Nature {\bf 412}, 620 (2001).
\bibitem{s5}  M. Beleggia and G. Pozzi, Phys. Rev. B{\bf 63}, 054507 (2001).

\bibitem{kinexp}     D.N. Basov et al, Phys. Rev. B{\bf 63}, 134514  (2001); 
 H. J. A. Molegraaf et al, Science {\bf 295}, 2239 (2002);
 A. F. Santander-Syro et al, Europhys. Lett., {\bf 62} 568 (2003);
A. Charnukha et al, 
Nature Communications {\bf 2}, 219 (2011);
C. Giannetti et al, Nature Comm. {\bf 2}, 353 (2011).

\bibitem{emf}  J.E. Hirsch,   J.  Sup. Nov. Mag. {\bf 23}, 309  (2010).



 \end{references}
 \end{document}